# Identification of Nonlinearity in Conductivity Equation via Dirichlet-to-Neumann Map[*]

Hyeonbae Kang[†]
School of Mathematical Sciences
Seoul National University
Seoul 151-747, Korea
hkang@math.snu.ac.kr

Gen Nakamura[‡]
Department of Mathematics
Hokkaido University
Sapporo 060-0810, Japan
gnaka@math.sci.hokudai.ac.jp

November 11, 2018

**Abstract**

We prove that the linear term and quadratic nonlinear term entering a nonlinear elliptic equation of divergence type can be uniquely identified by the Dirichlet to Neuman map. The unique identifiability is proved using the complex geometrical optics solutions and singular solutions.

## 1 Introduction

Let $\Omega$ be a bounded domain in $\mathbf{R}^n$ ($n \geq 2$) with a $C^2$ boundary $\partial\Omega$. We consider the following nonlinear boundary value problem:

(1.1) $$\begin{cases} \nabla \cdot C(x, \nabla u) = 0 & \text{in } \Omega, \\ u|_{\partial\Omega} = f. \end{cases}$$

The nonlinear function $C$ takes the following form:

(1.2) $\quad C(x, q) = \gamma q + Q(x, q) := \gamma q + P(x, q) + R(x, q), \quad x \in \overline{\Omega},\ q \in \mathbf{R}^n \text{ (or } \mathbf{C}^n\text{)}.$

---
[*]This work was completed while both authors were visiting the Mathematical Sciences Research Institute (MSRI). We would like to thank the institute for partial support and providing stimulating environment for the research.

[†]partly suppoted by KOSEF 98-0701-03-5 and 2000-1-10300-001-1.

[‡]partly supported by Grant-in-Aid for Scientific Research (C)(No.12640153).

Here $\gamma \in C^2(\overline{\Omega})$, and there exists $C_0 > 0$ such that

$$(1.3) \qquad \gamma(x) \geq C_0 \quad x \in \Omega.$$

$P(x,q)$ represents the quadratic nonlinearity:

$$(1.4) \qquad P(x,q) = \left( \sum_{1 \leq k \leq l \leq n} c^1_{kl}(x) q_k q_l, \cdots, \sum_{1 \leq k \leq l \leq n} c^n_{kl}(x) q_k q_l \right).$$

We suppose that $c^i_{kl} \in C^1(\overline{\Omega})$ are real valued functions and

$$(1.5) \qquad \|\gamma\|_{C^1(\overline{\Omega})}, \ \|c^i_{kl}\|_{C^1(\overline{\Omega})} \leq C_1 \quad (1 \leq i \leq n, 1 \leq k \leq l \leq n)$$

for some $C_1 > 0$. $R(x,q) \in C^2(\overline{\Omega} \times H)$ with $H := \{q \in \mathbf{R}^n \text{ or } \mathbf{C}^n \ ; \ |q| \leq h\}$ $(h > 0)$ represents the higher order nonlinearity, namely, it satisfies

$$(1.6) \qquad \begin{cases} |\partial_{x_j} R(x,q)| \leq C_2 |q|^3, \\ |\partial_{x_j} \partial_{q_k} R(x,q)| \leq C_2 |q|^2, \\ |\partial_{q_j} \partial_{q_k} R(x,q)| \leq C_2 |q|^1 \\ \qquad (1 \leq j, k \leq n, x \in \overline{\Omega}, q \in H) \end{cases}$$

for some $C_2 > 0$ independent of $x$ and $q$.

The following theorem is probably well-known. However, since we are not able to find a proper reference, we give a short proof of it based on the contraction mapping principle in the Appendix at the end of this paper. For the elasticity equation, a proof using the implicit function theorem can be found in [4].

**Theorem 1.1** *Let $n < p < \infty$. There exist $\epsilon$ and $\delta < h/2$ such that for any $f \in W^{2-1/p,p}(\partial\Omega)$ satisfying $\|f\|_{W^{2-1/p,p}(\partial\Omega)} < \epsilon$, (1.1) admits a unique solution $u$ such that $\|u\|_{W^{2,p}(\Omega)} < \delta$. Moreover there exists $C > 0$ independent of $f$ such that*

$$(1.7) \qquad \|u\|_{W^{2,p}(\Omega)} \leq C \|f\|_{W^{2-1/p,p}(\partial\Omega)}.$$

We define the Dirichlet-to-Neumann (DN) map $\Lambda_C(f)$ for $f$ with $\|f\|_{W^{2-1/p,p}(\partial\Omega)} < \epsilon$ to be

$$(1.8) \qquad \Lambda_C(f) := C(x, \nabla u)|_{\partial\Omega} \cdot \nu \in W^{1-1/p,p}(\partial\Omega)$$

where $u$ is the unique solution of (1.1) such that $\|u\|_{W^{2,p}(\Omega)} < \delta$.

In this paper we consider the inverse boundary value problem to identify $C$ by means of the DN map $\Lambda_C$. We are particularly interested in finding the linear term $\gamma$ and the second order nonlinearity $P(x,q)$.

This inverse problem has interest in it's own right as the one to find conductivity distribution when the conductivity varies depending on the currents. It may also considered as a simplified model for the nonlinear elasticity equation. In elasticity $\gamma(x)$ corresponds to the Lamé moduli and $c_{kl}^l(x)$ can be thought as the third order elasticity tensor. In acousto-elasticity, this higher elasticity tensor is important [5].

We obtain the following unique identifiability theorem for $\gamma$ and $P$.

**Theorem 1.2** Suppose that $n \geq 3$. Let $C^{(j)}(x,q) = \gamma_j q + Q^{(j)}$, $j = 1, 2$, where $Q^{(j)} = P^{(j)}(x,q) + R^{(j)}(x,q)$,

$$P^{(j)}(x,q) = \left( \sum_{1 \leq k \leq l \leq n} c_{kl}^{(j)1}(x) q_k q_l, \cdots, \sum_{1 \leq k \leq l \leq n} c_{kl}^{(j)n}(x) q_k q_l \right),$$

and $R^{(j)}$ satisfy (1.3), (1.5), and (1.6), respectively, for the same constants $C_0, C_1, C_2$ and $h$. If

(1.9) $\quad \Lambda_{C^{(1)}}(f) = \Lambda_{C^{(2)}}(f) \quad$ for all complex-valued small $f \in W^{2-1/p,p}(\partial\Omega)$,

then

(1.10) $\qquad \gamma_1(x) = \gamma_2(x) \quad$ and $\quad c_{kl}^{(1)j}(x) = c_{kl}^{(2)j}(x), \quad x \in \Omega$

for $1 \leq j \leq n$ and $1 \leq k \leq l \leq n$.

The dimensional restriction $n \geq 3$ is imposed since we are using the complex geometrical optics solutions of Sylvester-Uhlmann [13] to prove the Theorem.

In most models of nonlinear elasticity, the higher order tensors do not depend on $x$. So we consider the same inverse problem when the coefficients $c_{kl}^j$ are constants. In this case we obtain the following uniqueness theorem including the two dimensions. One thing to be noted is the condition (1.9) (and (1.11) below). Since the equation considered in this paper is nonlinear, $\Lambda_{C^{(1)}}(f) = \Lambda_{C^{(2)}}(f)$ for all real-valued small $f \in W^{2-1/p,p}(\partial\Omega)$ does not imply the same for complex-valued small $f \in W^{2-1/p,p}(\partial\Omega)$. Therefore, in the situation where the data $\Lambda_C(f)$ for only real-valued small $f$ are available, Theorem 1.2 may not be applied. The following theorem uses only real-valued Dirichlet data.

**Theorem 1.3** Suppose that $n \geq 2$. Let $C^{(j)}(x,q)$ as in Theorem 1.2 except that the coefficients $c_{kl}^{(j)i}$ are constants. If

(1.11) $\quad \Lambda_{C^{(1)}}(f) = \Lambda_{C^{(2)}}(f) \quad$ for all real-valued small $f \in W^{2-1/p,p}(\partial\Omega)$,

then

(1.12) $\qquad \gamma_1(x) = \gamma_2(x) \quad$ and $\quad c_{kl}^{(1)j} = c_{kl}^{(2)j}$

for $1 \leq j \leq n$ and $1 \leq k \leq l \leq n$.

There have been some related works on the identification of nonlinear terms entering partial differential equations. Sun proved the global uniqueness for the nonlinear conductivity $\gamma(x, u)$ when the Dirichlet data are complex valued and small [11]. Nakamura and Sun considered the similar problem for nonlinear elasticity model of St.Venant-Kirchhoff [10]. They proved that even in the presence of nonlinearity, the linear term can be identified uniquely by means of DN map. Sun and Uhlmann proved the global uniqueness up to diffeomorphism fixing the boundary for the two dimensional nonlinear anisotropic conductivity $A(x, u)$, and also the same result for the 3 dimensional analytic conductivity [12]. Quite recently, Isakov proved the unique identifiability of the nonlinearity $c(u, p)$ entering the quasilinear elliptic equation $\Delta u + c(u, \nabla u) = 0$ (and corresponding parabolic equation) by means of DN map on $\partial \Omega$ [7]. G. Uhlmann informed us that Hervas and Sun proved a result related to ours in the two dimensional case [6]. They proved that the constant coefficients nonlinear terms with extra symmetry can be identified from the DN map defined for complex valued small Dirichlet data.

Theorem 1.2 and Theorem 1.3 are proved by investigating the first and second terms in the asymptotic expansion of $\Lambda_C$ near 0. The first term is $\Lambda_\gamma$, the DN map corresponding to the conductivity $\gamma$. This fact was also observed in [10] and [7]. We derive certain cubic relation from the second term. This is included in Section 2. Then using the complex geometrical optics solutions of Sylvester-Uhlmann [13], we prove Theorem 1.2 in Section 3. We prove Theorem 1.3 using the singular solutions of Alessandrini [2]. Since the coefficients to be determined are constants, boundary determination is sufficient. This is the reason why the singular solutions are effectively used. The proof is included in Section 4. Appendix is to prove Theorem 1.1.

## 2 Asymptotics of DN map

Throughout this paper $\| \ \|_{k,p}$ denotes the $W^{k,p}$-norm on $\Omega$, and $\| \ \|_p = \| \ \|_{0,p}$. Also, $C > 0$ denotes the general constant in estimate independent of the functions being estimated. We will use the following estimates repeatedly: if $k \geq 1$, $p > n$, and $n = 2, 3$, then

(2.1) $$\|uv\|_{k,p} \leq C \|u\|_{k,p} \|v\|_{k,p}.$$

This inequality holds by the Sobolev embedding theorem (see [1]). We also use the chain rule for the derivative of the composition $R(x, \nabla u(x))$ for $u \in W^{2,p}(\Omega)$ and its estimate given in Chapter II, section 3 of [14].

Suppose that $f \in W^{2-1/p,p}(\partial\Omega)$, $p > n$, and let $t \in \mathbf{R}^1$ be a small parameter. Let $u^{(t)}$ be the solution of

$$\begin{cases} \nabla \cdot C(x, \nabla u) = 0 & \text{in } \Omega, \\ u|_{\partial\Omega} = tf \end{cases}$$

such that $\|u^{(t)}\|_{2,p} \le \delta$. Then there exists a constant $C > 0$ such that

(2.2) $$\|u^{(t)}\|_{2,p} \le Ct\|f\|_{W^{2-1/p,p}(\partial\Omega)}.$$

Let $u_1 = u_1(f) \in W^{2,p}(\Omega)$ be the solution of

$$\begin{cases} \nabla \cdot (\gamma \nabla u_1) = 0 & \text{in } \Omega, \\ u_1|_{\partial\Omega} = f. \end{cases}$$

Then by the regularity of the Dirichlet problem for elliptic equations, there exist a constant $C > 0$ such that we have

(2.3) $$\|u_1(f)\|_{2,p} \le C\|f\|_{W^{2-1/p,p}(\partial\Omega)}$$

for any $f \in W^{2-1/p,p}(\partial\Omega)$. Let $u_2$ be the solution of

(2.4) $$\begin{cases} \nabla \cdot (\gamma \nabla u_2) = -\nabla \cdot P(x, \nabla u_1) & \text{in } \Omega, \\ u_2|_{\partial\Omega} = 0. \end{cases}$$

It then follows from the regularity of elliptic equations, (2.1), and (2.3) that

$$\begin{aligned} \|u_2\|_{2,p} &\le C\|\nabla \cdot P(x, \nabla u_1)\|_p \\ &\le C\|\nabla u_1\|_{1,p}^2 \\ &\le C\|f\|_{W^{2-1/p,p}(\partial\Omega)}^2. \end{aligned}$$

We now define $v^{(t)}$ by

(2.5) $$u^{(t)} = t(u_1 + tv^{(t)}).$$

Then, we have from (2.2) and (2.3)

$$\begin{aligned} t^2\|v^{(t)}\|_{2,p} &\le \|u^{(t)}\|_{2,p} + t\|u_1\|_{2,p} \\ &\le Ct\|f\|_{W^{2-1/p,p}(\partial\Omega)}. \end{aligned}$$

**Lemma 2.1** *There exist $t_0$ and $C_f$ depending on $f$, not on $t$, such that for all $t < t_0$,*

(2.6) $$\|v^{(t)} - u_2\|_{2,p} \le C_f t.$$

*In particular, we have*

(2.7) $$\|v^{(t)}\|_{2,p} \le C_f.$$

*Proof.* Put $w^{(t)} := v^{(t)} - u_2$. Then a straight-forward computation shows that
$$\begin{cases} \nabla \cdot \gamma \nabla w^{(t)} = -t g^{(t)} & \text{in } \Omega, \\ w^{(t)}|_{\partial \Omega} = 0 \end{cases}$$
where
$$g^{(t)} = t^{-1} \nabla \cdot \left[ P(x, \nabla u_1 + t \nabla v^{(t)}) - P(x, \nabla u_1) \right] + t^{-3} \nabla \cdot R(x, t \nabla u_1 + t^2 \nabla v^{(t)}).$$
It follows from (1.4), (1.6), and (2.1) that
$$\|g^{(t)}\|_p \leq C(\|u_1\|_{2,p} \|v^{(t)}\|_{2,p} + t \|v^{(t)}\|_{2,p}^2 + \|u_1\|_{2,p}^3 + t^3 \|v^{(t)}\|_{2,p}^3).$$
It then follows from (2) that
$$\|v^{(t)} - u_2\|_{2,p} = \|w^{(t)}\|_{2,p} \leq Ct \|g^{(t)}\|_p \\ \leq C(t \|v^{(t)}\|_{2,p} + t).$$
Thus, if $t_0$ is so small that $Ct_0 < \frac{1}{2}$, then by (2) we obtain
$$\|v^{(t)}\|_{2,p} \leq 2(\|u_2\|_{2,p} + 1) \leq C.$$
Then (2.6) follows from (2). This completes the proof. □

**Lemma 2.2** *For a given $f \in W^{2-1/p,p}(\partial \Omega)$, we have*

(1) $\lim_{t \to 0} \frac{1}{t} [\Lambda_C(tf) - t \Lambda_\gamma(f)] = 0$,

(2) $\lim_{t \to 0} \frac{1}{t^2} [\Lambda_C(tf) - t \Lambda_\gamma(f)] = \nu \cdot [\gamma \nabla u_2 + P(x, \nabla u_1)]$.

*The convergence is in the topology of $W^{1-1/p,p}(\partial \Omega)$.*

*Proof.* From (2.5) we have
$$\Lambda_C(tf) - t \Lambda_\gamma(f) = \nu \cdot [C(x, \nabla u^{(t)}) - t \gamma \nabla u_1] \\ = \nu \cdot [t^2 \gamma \nabla v^{(t)} + Q(x, \nabla u^{(t)})].$$
It follows from the trace theorem [1], (2.2) and (2.7) that
$$\|t^{-1}(\Lambda_C(tf) - t \Lambda_\gamma(f))\|_{W^{1-1/p,p}(\partial \Omega)} \leq \|t \gamma \nabla v^{(t)} + t^{-1} Q(x, \nabla u^{(t)})\|_{1,p} \\ \leq Ct.$$
This proves (1).

Since
$$Q(x, \nabla u^{(t)}) = t^2 P(x, \nabla u_1) + t^3 O(|\nabla u_1||\nabla v^{(t)}| + |\nabla v^{(t)}|^2 + |\nabla u_1|^3 + |\nabla v^{(t)}|^3),$$
(2) follows from (2.6). This completes the proof. □

# 3 Proof of Theorem 1.2

Lemma 2.2 says that we can recover $\Lambda_\gamma$ and $\nu \cdot [\gamma \nabla u_2 + P(x, \nabla u_1)]$ on $\partial \Omega$ from $\Lambda_C$.

If $\Lambda_{C^{(1)}} = \Lambda_{C^{(2)}}$, then $\Lambda_{\gamma_1} = \Lambda_{\gamma_2}$. It then follows from well-known results ([3], [8, 9], [13]) that $\gamma_1 = \gamma_2$ in $\Omega$. Put $\gamma = \gamma_1 = \gamma_2$. By Lemma 2.2 (2), we have

$$(3.1) \quad \nu \cdot [\gamma \nabla u_2^{(1)} + P^{(1)}(x, \nabla u_1)] = \nu \cdot [\gamma \nabla u_2^{(2)} + P^{(2)}(x, \nabla u_1)] \quad \text{on } \partial \Omega$$

where $u_1 \in W^{2,p}(\Omega)$ is a solution of $\nabla \cdot (\gamma \nabla u_1) = 0$ in $\Omega$, and $u_2^{(j)}$ is the solution of (2.4) when $P = P^{(j)}$.

Let $v \in W^{2,p}(\Omega)$ be a solution of $\nabla \cdot (\gamma \nabla v) = 0$ in $\Omega$ with $v|_{\partial \Omega} = g$. Since $u_2^{(j)}|_{\partial \Omega} = 0$, it follows from the divergence theorem that

$$\int_{\partial \Omega} \nu \cdot [\gamma \nabla u_2^{(j)} + P^{(j)}(x, \nabla u_1)] g d\sigma = \int_\Omega [\gamma \nabla u_2^{(j)} + P^{(j)}(x, \nabla u_1)] \cdot \nabla v dx$$
$$= \int_\Omega P^{(j)}(x, \nabla u_1) \cdot \nabla v dx.$$

We thus have

$$(3.2) \quad \int_\Omega P^{(1)}(x, \nabla u_1) \cdot \nabla v dx = \int_\Omega P^{(2)}(x, \nabla u_1) \cdot \nabla v dx,$$

in other words,

$$(3.3) \quad \sum_{i=1}^n \sum_{1 \leq j \leq l \leq n} \int_\Omega c_{jl}^{(1)i}(x) \frac{\partial u_1}{\partial x_j} \frac{\partial u_1}{\partial x_l} \frac{\partial v}{\partial x_i} dx = \sum_{i=1}^n \sum_{1 \leq j \leq l \leq n} \int_\Omega c_{jl}^{(2)i}(x) \frac{\partial u_1}{\partial x_j} \frac{\partial u_1}{\partial x_l} \frac{\partial v}{\partial x_i} dx$$

for all $u_1, v \in W^{2,p}(\Omega)$, solutions of $\nabla \cdot \gamma \nabla u = 0$ in $\Omega$. Therefore, it suffices to prove that if

$$(3.4) \quad \sum_{i=1}^n \sum_{1 \leq j \leq l \leq n} \int_\Omega c_{jl}^i(x) \frac{\partial u_1}{\partial x_j} \frac{\partial u_1}{\partial x_l} \frac{\partial v}{\partial x_i} dx = 0$$

for all $u_1, v \in W^{2,p}(\Omega)$, solutions of $\nabla \cdot \gamma \nabla u = 0$, then

$$(3.5) \quad c_{jl}^i \equiv 0, \quad i = 1, \cdots, n, \ 1 \leq j \leq l \leq n.$$

Let $u_2 \in W^{2,p}(\Omega)$ be another solution of $\nabla \cdot \gamma \nabla u = 0$. Then $u_1 + u_2$ is also a solution of the equation, and hence we obtain by polarizing (3.4) that

$$(3.6) \quad \sum_{i=1}^n \sum_{1 \leq j \leq l \leq n} \int_\Omega c_{jl}^i(x) \left( \frac{\partial u_1}{\partial x_j} \frac{\partial u_2}{\partial x_l} + \frac{\partial u_1}{\partial x_l} \frac{\partial u_2}{\partial x_j} \right) \frac{\partial v}{\partial x_i} dx = 0$$

for all $u_1, u_2, v \in W^{2,p}(\Omega)$, solutions of $\nabla \cdot \gamma \nabla u = 0$. The arguments we made so far are true for $n \geq 2$.

Let $n = 3$ and use complex geometrical optics solutions. Let $k \in \mathbf{R}^3$ and choose unit vectors $\xi$ and $\eta$ in $\mathbf{R}^3$ so that

(3.7) $$k \cdot \xi = k \cdot \eta = \xi \cdot \eta = 0.$$

Then choose $t, s > 0$ so that

(3.8) $$t^2 = \frac{|k|^2}{4} + s^2.$$

Define $\rho^{(1)}, \rho^{(2)} \in \mathbf{C}^3$ by

(3.9) $$\rho^{(1)} = t\eta + i(\frac{k}{2} + s\xi), \quad \rho^{(2)} = -t\eta + i(\frac{k}{2} - s\xi)$$

Then by the fundamental work of Sylvester-Uhlmann [13] there exist solutions $u_j$, $j = 1, 2$, of $\nabla \cdot \gamma \nabla u = 0$ of the form

(3.10) $$u_j(x) = \gamma^{-1/2} e^{\rho^{(j)} \cdot x}(1 + \psi_j(x, \rho^{(j)})), \quad x \in \Omega$$

where $\psi_j$ satisfies

(3.11) $$\|\psi_j\|_{L^\infty(\Omega)} \leq \frac{C}{|s|} \quad \text{and} \quad \|\nabla \psi_j\|_{L^\infty(\Omega)} \leq C$$

for some $C$ independent of $t$. We apply these solutions $u_j$ to (3.6) and obtain

(3.12)
$$\sum_{i=1}^{3} \sum_{1 \leq j \leq l \leq 3} \int_\Omega \gamma(x)^{-1} c_{jl}^i(x) \frac{\partial v}{\partial x_i}(x) e^{ik \cdot x} \times$$
$$\left[ (\gamma^{1/2} \partial_j \gamma^{-1/2}(1+\psi_1) + \rho_j^{(1)}(1+\psi_1) + \partial_j \psi_1) \times \right.$$
$$(\gamma^{1/2} \partial_l \gamma^{-1/2}(1+\psi_2) + \rho_l^{(2)}(1+\psi_2) + \partial_l \psi_2)$$
$$+ (\gamma^{1/2} \partial_l \gamma^{-1/2}(1+\psi_1) + \rho_l^{(1)}(1+\psi_1) + \partial_l \psi_1) \times$$
$$\left. (\gamma^{1/2} \partial_j \gamma^{-1/2}(1+\psi_2 + \rho_j^{(2)}(1+\psi_2) + \partial_j \psi_2) \right] dx$$
$$= 0.$$

Here $\partial_j = \partial/\partial x_j$. Set $\zeta := \eta + i\xi$ and note that

$$\lim_{s \to \infty} \frac{\rho^{(1)}}{s} = \zeta, \quad \lim_{s \to \infty} \frac{\rho^{(2)}}{s} = -\zeta.$$

Therefore, by dividing the both sides of (3.12) by $t$, taking the limit $s \to \infty$, and using (3.11), we get

$$(3.13) \qquad \int_\Omega \sum_{i=1}^3 \sum_{1 \leq j \leq l \leq 3} \zeta_j \zeta_l c_{jl}^i(x) \gamma(x)^{-1} \frac{\partial v}{\partial x_i}(x) e^{ik \cdot x} dx = 0.$$

Since (3.13) holds for all $k \in \mathbf{R}^3$, we have

$$(3.14) \qquad \sum_{i=1}^3 \left[ \sum_{1 \leq j \leq l \leq 3} \zeta_j \zeta_l c_{jl}^i(x) \right] \frac{\partial v}{\partial x_i}(x) = 0, \quad x \in \Omega.$$

We need the following lemma the proof of which will be given at the end of this section.

**Lemma 3.1** *Suppose $n \geq 2$. There exist solutions $v_j \in W^{2,p}(\Omega)$, $j = 1, \cdots, n$, of $\nabla \cdot \gamma \nabla v_j = 0$ in $\Omega$ such that*

$$(3.15) \qquad \det\left(\frac{\partial v_j}{\partial x_i}(x)\right) \neq 0, \quad \text{for almost all } x \in \Omega.$$

Lemma 3.1 and (3.14) yield

$$(3.16) \qquad \sum_{1 \leq j \leq l \leq n} \zeta_j \zeta_l c_{jl}^i(x) = 0, \quad x \in \Omega, \ i = 1, 2, 3.$$

Note that (3.16) holds for all $\zeta$ in the set $\mathcal{V} := \{\zeta \in \mathbf{C}^3 | \ \zeta \cdot \zeta = 0, \ |\zeta| = \sqrt{2}\}$. If $\zeta = (0, z, \sqrt{-1}z)$ for $z \in \mathbf{C}$ with $|z| = 1$, then $\zeta \in \mathcal{V}$. With this $\zeta$, (3.16) becomes

$$(3.17) \qquad c_{22}^i(x) - c_{33}^i(x) + \sqrt{-1} c_{23}^i(x) = 0, \quad x \in \Omega.$$

Since $c_{jl}^i$ are real, we have $c_{22}^i(x) = c_{33}^i(x)$ and $c_{23}^i = 0$. By substituting $\zeta = (z, 0, \sqrt{-1}z)$ and $\zeta = (z, \sqrt{-1}z, 0)$ in order into (3.16), we obtain that

$$(3.18) \qquad c_{11}^i = c_{22}^i = c_{33}^i \quad \text{and} \quad c_{jl}^i = 0 \ (j \neq l).$$

Because of (3.18), (3.6) now takes the form

$$(3.19) \qquad \sum_{i=1}^3 \sum_{j=1}^3 \int_\Omega c_{jj}^i(x) \frac{\partial u_1}{\partial x_j} \frac{\partial u_2}{\partial x_j} \frac{\partial v}{\partial x_i} dx = 0$$

for all $u, u_2, v$. By the same argument as above (using $u_2$ and $v$), we can conclude that

$$(3.20) \qquad \sum_{j=1}^3 \left[ \zeta_j \sum_{i=1}^3 \zeta_i c_{jj}^i(x) \right] \frac{\partial u_1}{\partial x_j}(x) = 0$$

for all $u_1$ and $\zeta \in \mathcal{V}$. It then follows from Lemma 3.1 that

$$\zeta_j \sum_{i=1}^{3} \zeta_i c^i_{jj}(x) = 0, \quad j = 1, 2, 3, \tag{3.21}$$

and we conclude that $c^i_{jj} = 0$, $i, j = 1, 2, 3$. The proof is complete. $\square$

*Proof of Lemma 3.1.* We assume that $n = 3$. The same proof works for the two dimensional case since complex geometrical optics solutions exist in two dimensions. Choose $\zeta^{(1)}, \zeta^{(2)}, \zeta^{(3)} \in \mathcal{V}$ so that they are linearly independent over $\mathbf{C}$, e.g., $\zeta^{(1)} = (1, \sqrt{-1}, 0)$, $\zeta^{(2)} = (1, 0, \sqrt{-1})$, $\zeta^{(3)} = (0, 1, \sqrt{-1})$. Let

$$v_j(x) = \gamma^{-1/2} e^{t\zeta^{(j)} \cdot x}(1 + \psi_j(x, t\zeta^{(j)})), \quad x \in \Omega, \ j = 1, 2, 3$$

as before. Then, we have

$$\det \begin{pmatrix} \nabla v_1 \\ \nabla v_2 \\ \nabla v_3 \end{pmatrix} = \gamma^{-3/2} e^{t(\zeta^{(1)} + \zeta^{(2)} + \zeta^{(3)}) \cdot x} t^3 \det \begin{pmatrix} \zeta^{(1)}(1 + \psi_1) + O(t^{-1}) \\ \zeta^{(2)}(1 + \psi_2) + O(t^{-1}) \\ \zeta^{(3)}(1 + \psi_3) + O(t^{-1}) \end{pmatrix}.$$

It then follows from (3.11) that if $t$ is sufficiently large, then

$$\det \begin{pmatrix} \nabla v_1(x) \\ \nabla v_2(x) \\ \nabla v_3(x) \end{pmatrix} \neq 0.$$

This finishes the proof. $\square$

## 4 Proof of Theorem 1.3

We need to prove that if

$$\sum_{i=1}^{n} \sum_{1 \leq k \leq l \leq n} c^i_{kl} \int_\Omega \frac{\partial u}{\partial x_k} \frac{\partial u}{\partial x_l} \frac{\partial v}{\partial x_i} dx = 0 \tag{4.1}$$

for all $u, v \in W^{2,p}(\Omega)$ real solutions of $\nabla \cdot \gamma \nabla u = 0$, then

$$c^i_{kl} = 0, \quad i = 1, \cdots, n, \ 1 \leq k \leq l \leq n. \tag{4.2}$$

Since (4.1) holds only for real-valued solutions $u, v$, we can not use the complex geometrical optics solutions. However, since the coefficients $c^i_{kl}$ are assumed to be constants, we can use instead the singular solutions of Alessandrini. We only deal with the three dimensional case. Two dimensional case can be proved in the same way.

Fix $j$ and $s \leq t$. We will show that $c_{st}^j = 0$. Let $e_k$, $k = 1, 2, 3$, is the standard basis of $\mathbf{R}^3$. Let $\alpha > 0$, $\beta > 0$, $\alpha^2 + \beta^2 = 1$. Define an orthonormal frame $N, T_1, T_2$ by $N = \alpha e_s + \beta e_t$, $T_2 = \beta e_s - \alpha e_t$, and $T_1 = e_k$ where $k \neq s, t$ if $s \neq t$. In the case $s = t$, let $N = e_s$, and $T_1$ and $T_2$ be $e_k$ and $e_l$ where $k \neq l \neq s \neq k$. For $x \in \partial\Omega$, $\nu(x)$ denote the unit outward normal to $\partial\Omega$ at $x$. Choose $x_0 \in \partial\Omega$ so that $\nu(x_0) = N$. By translation if necessary, we may assume that $x_0 = 0$.

Choose a solution $v \in C^2(\overline{\Omega})$ of $\nabla \cdot \gamma \nabla v = 0$ so that $\partial_i v(0) = \delta_{ij}$ ($i = 1, 2, 3$). It can be proved easily that such a function exists. Then

$$(4.3) \qquad \partial_i v(x) = \delta_{ij} + O(|x|), \quad x \to 0.$$

Let $U$ be an open neighborhood of $0$. Then we have from (4.1) and (4.3) that

$$(4.4) \qquad \left| \sum_{1 \leq k \leq l \leq n} c_{kl}^j \int_{\Omega \cap U} \partial_k u \partial_l u \, dx \right| \leq C \int_\Omega |x| |\nabla u|^2 dx$$

for some $C > 0$.

By [2], for every $\epsilon > 0$, there exists a solution $u \in C^2(\overline{\Omega})$ such that

$$u(x) = \frac{1}{|x - \epsilon N|} + w(x)$$

where

$$(4.5) \qquad |\nabla w(x)| \leq C|x - \epsilon N|^{-2+\alpha}$$

for some $\alpha > 0$ ($\alpha < 1$) and $C$ independent of $\epsilon$. In two dimensions we can use $\log|x - \epsilon N|$. Put $\Gamma_\epsilon(x) := |x - \epsilon N|^{-1}$ to make notations short. Substituting this solution $u$ to (4.4), we get

$$(4.6) \qquad \left| \sum_{k \leq l} c_{kl}^j \int_{\Omega \cap U} \partial_k \Gamma_\epsilon \partial_l \Gamma_\epsilon dx \right| \leq C \int_\Omega |x - \epsilon N|^{-4+\alpha} dx.$$

For convenience, put $y := x - \epsilon N$. From (4.6), we have

$$(4.7) \qquad \left| \sum_{k \leq l} c_{kl}^j \int_{\Omega \cap U} \frac{y_k y_l}{|y|^6} dx \right| \leq C \int_\Omega |x - \epsilon N|^{-4+\alpha} dx.$$

Suppose $\partial\Omega$ near $0$ is given by

$$\partial\Omega \cap U = \{\, \xi_1 T_1 + \xi_2 T_2 - \varphi(\xi_1, \xi_2) N \mid |\xi_1|^2 + |\xi_2|^2 < \delta^2 \,\}$$

where $\varphi \in C^2$, $\varphi(0) = 0$, and $\nabla\varphi(0) = 0$, for a fixed small $\delta > 0$. Let $D := \{\, \xi_1 T_1 + \xi_2 T_2 - \xi_3 N \mid |\xi_1|^2 + |\xi_2|^2 < \delta^2,\ 0 < \xi_3 < \delta \,\}$ and define $\Phi : D \to \Omega \cap U$ by

$$\Phi(\xi_1 T_1 + \xi_2 T_2 - \xi_3 N) := \xi_1 T_1 + \xi_2 T_2 - (\xi_3 + \varphi(\xi_1, \xi_2))N.$$

By shrinking $U$ if necessary, we may assume that $\Phi(D) = \Omega \cap U$. Then we have after a coordinate change, (4.7) takes the form

(4.8) $$\left| \sum_{k \leq l} c_{kl}^j \int_D \frac{z_k z_l}{|z|^6} d\xi \right| \leq C \int_\Omega |x - \epsilon N|^{-4+\alpha} dx.$$

where $z = \Phi(\xi) - \epsilon N$, $\xi = \xi_1 T_1 + \xi_2 T_2 - \xi_3 N \in D$. We now scale $\eta \to \xi = \epsilon \eta$ to obtain

(4.9) $$\int_D \frac{z_k z_l}{|z|^6} d\xi = \epsilon^{-1} \int_{|\eta_1|^2 + |\eta_2|^2 < (\delta/\epsilon)^2} \int_{\eta_3 = 0}^{\delta/\epsilon} \frac{\zeta_k \zeta_l}{|\zeta|^6} d\eta$$

where $\zeta = \eta_1 T_1 + \eta_2 T_2 - (\eta_3 + \epsilon^{-1}\varphi(\epsilon\eta_1, \epsilon\eta_2) + 1)N$. It is easy to prove, by scaling, that

(4.10) $$\lim_{\epsilon \to 0} \epsilon \int_\Omega |x - \epsilon N|^{-4+\alpha} dx = 0.$$

Let $\eta' := (\eta_1, \eta_2)$. Since

$$\epsilon^{-1}|\varphi(\epsilon\eta')| \leq C\epsilon|\eta'|^2 \leq C\delta|\eta'|, \quad \text{for } |\eta'| < \delta/\epsilon,$$

one can see that

$$\frac{\zeta_k \zeta_l}{|\zeta|^6} \leq C \frac{1}{|\eta|^4 + 1}$$

for some $C$ independent of $\epsilon$. Therefore, we obtain from the dominated convergence theorem

(4.11) $$\lim_{\epsilon \to 0} \int_{|\eta'| < \delta/\epsilon} \int_0^{\delta/\epsilon} \frac{\zeta_k \zeta_l}{|\zeta|^6} d\eta = \int_{\mathbf{R}_+^3} \frac{w_k w_l}{|w|^6} d\eta$$

where $w = \eta_1 T_1 + \eta_2 T_2 - (\eta_3 + 1)N$ and $\mathbf{R}_+^3$ is the upper half space. Let

$$I_{kl}^{st} := \int_{\mathbf{R}_+^3} \frac{w_k w_l}{|w|^6} d\eta.$$

It follows from (4.8), (4.9), (4.10) and (4.11) that

(4.12) $$\sum_{k \leq l} c_{kl}^j I_{kl}^{st} = 0.$$

Suppose that $s = t$. In this case $N$ was chosen to be $N = e_s$. Thus $w_k = \eta_k$ if $k \neq s$ and $w_s = -(\eta_s + 1)$. Therefore easy computations show that

$$I_{kl}^{ss} = \begin{cases} 0 & \text{if } k \neq l, \\ \dfrac{\pi}{4} & \text{if } k = l \neq s, \\ \dfrac{\pi}{2} & \text{if } k = l = s. \end{cases}$$

Thus by (4.12) we get

(4.13) $$2c_{ss}^j + \sum_{k \neq s} c_{kk}^j = 0, \quad s = 1, 2, 3.$$

Hence we conclude that

(4.14) $$c_{kk}^j = 0, \quad j, k = 1, 2, 3,$$

and (4.12) takes the form

(4.15) $$\sum_{k<l} c_{kl}^j I_{kl}^{st} = 0.$$

Suppose that $s < t$. If $k \neq s, t$, then $T_1 = e_k$ and hence $w_k = \eta_1$, $w_s = \beta\eta_2 - \alpha(\eta_3 + 1)$, and $w_t = -\alpha\eta_2 - \beta(\eta_3 + 1)$. Again by simple computations one can see that

$$I_{kl}^{st} = \begin{cases} 0 & \text{if } (k,l) \neq (s,t), \ k < l, \\ \alpha\beta\dfrac{\pi}{4} & \text{if } (k,l) = (s,t). \end{cases}$$

It then follows from (4.15) that for $s < t$

(4.16) $$c_{st}^j = 0, \quad j = 1, 2, 3.$$

By (4.14) and (4.16), the proof is complete. $\square$

## 5 Appendix- Proof of Theorem 1.1

Replace $R(x,q)$ by $\tilde{R}(x,q) := \chi(q)R(x,q) + (1 - \chi(q))|q|^3$ with $\chi \in C_0^\infty(\mathbf{R}^n \text{ or } \mathbf{C}^n)$ satisfying $0 \leq \chi(q) \leq 1$, $\chi(q) = 1$ ($|q| \leq h/2$), $0$ ($|q| \geq h$). Here, for the complex vector $q \in \mathbf{C}^n$ is identified with $(\text{Re}q, \text{Im}q) \in \mathbf{R}^{2n}$ and $\chi(q)$ is considered as a function of variables $\text{Re}q$, $\text{Im}q$. Then, (1.6) holds for any $q \in \mathbf{R}^n$ or $\mathbf{C}^n$ with some new $C_2$ determined by $h$, $\chi(q)$ and the old $C_2$. Suppose that $\|f\|_{W^{2-1/p,p}(\partial\Omega)} \leq \epsilon$ ($\epsilon$ to be determined later). Let $u_0$ be the solution of

(5.1) $$\begin{cases} \nabla \cdot \gamma \nabla u_0 = 0, & \text{in } \Omega, \\ u_0|_{\partial\Omega} = f. \end{cases}$$

If $u$ is a solution of (1.1), then $v$ defined by $u = u_0 + v$ satisfies

(5.2)
$$\begin{cases} \nabla \cdot \gamma \nabla v + \nabla \cdot Q(x, \nabla u_0 + \nabla v) = 0, & \text{in } \Omega, \\ v|_{\partial \Omega} = 0. \end{cases}$$

Let $L_\gamma^{-1}$ be the solution operator for the problem

(5.3)
$$\begin{cases} \nabla \cdot \gamma \nabla u = g \in L^p(\Omega), & \text{in } \Omega, \\ u|_{\partial \Omega} = 0, \end{cases}$$

namely, $L_\gamma^{-1} g$ is the solution of (5.3). Then $L_\gamma^{-1}$ is linear and there exists $C > 0$ independent of $g$ such that
$$\|L_\gamma^{-1} g\|_{2,p} \leq C \|g\|_p.$$

Let $X := \{\varphi \in W^{2,p}(\Omega) \mid \varphi|_{\partial \Omega} = 0, \ \|\varphi\|_{2,p} \leq \delta \}$ ($\delta$ to be determined later) and define an operator $A$ on $X$ by
$$A(\varphi)(x) := L_\gamma^{-1} \nabla \cdot Q(x, \nabla u_0 + \nabla \varphi).$$

Then, it follows from (1.4) and (1.6) that

(5.4)
$$\begin{aligned} \|A(\varphi)\|_{2,p} &\leq C \|Q(\nabla u_0 + \nabla \varphi)\|_{1,p} \\ &\leq C(\|u_0\|_{2,p}^2 + \|\varphi\|_{2,p}^2 + \|u_0\|_{2,p}^3 + \|\varphi\|_{2,p}^3) \\ &\leq C(\epsilon^2 + \delta^2 + \epsilon^3 + \delta^3), \end{aligned}$$

and

(5.5)
$$\begin{aligned} \|A(\varphi_1) - A(\varphi_2)\|_{2,p} &\leq C \|Q(x, \nabla u_0 + \nabla \varphi_1) - Q(x, \nabla u_0 + \nabla \varphi_2)\|_{1,p} \\ &\leq C(\|u_0 + \theta \varphi_1 + (1-\theta)\varphi_2\|_{2,p} + \|u_0 + \theta \varphi_1 + (1-\theta)\varphi_2\|_{2,p}^2) \\ &\quad \|\nabla(\varphi_1 - \varphi_2)\|_{1,p} \\ &\leq C(\epsilon + \delta + \epsilon^2 + \delta^2) \|\varphi_1 - \varphi_2\|_{2,p} \end{aligned}$$

for some $0 < \theta < 1$. Therefore, if $C(\epsilon^2 + \delta^2 + \epsilon^3 + \delta^3) \leq \delta$ and $C(\epsilon + \delta + \epsilon^2 + \delta^2) < 1$, then $A$ is a contraction on $X$.

By the contraction mapping principle, there exists a unique fixed point of $A$, say $v$. Then $v$ is the solution of (5.3). The estimate (1.7) follows from the equation and regularity of the linear elliptic equation.

Finally, we take $\epsilon$ and $\delta$ more smaller if necessary so that they satisfy $C_3 \epsilon + \delta \leq h/2$. Then, $u = u_0 + v$ is the solution of (1.1).

*Acknowledgement.* The authors thank Prof. Kubota Koji for a useful comment.